\documentclass{Interspeech2024}

\interspeechcameraready

\title{Large Language Model-based FMRI Encoding of Language Functions for Subjects with Neurocognitive Disorder}

\usepackage{siunitx}
\usepackage{multirow}
\usepackage{amsmath}
\usepackage{cite}

\name[affiliation={1*}]{Yuejiao}{Wang}
\name[affiliation={2,3*}]{Xianmin}{Gong}
\name[affiliation={1}]{Lingwei}{Meng}
\name[affiliation={1}]{Xixin}{Wu}
\name[affiliation={1,2}]{Helen}{Meng}

\address{
  $^1$Dept. of Systems Engineering \& Engineering Management, The Chinese University of Hong Kong \\
  $^2$Stanley Ho Big Data Decision Analytics Research Centre, The Chinese University of Hong Kong\\
  $^3$Department of Psychology, The Chinese University of Hong Kong
 }

\email{\{wangy, lmeng, wuxx, hmmeng\}@se.cuhk.edu.hk xianmingong@cuhk.edu.hk $^*$Equal contributions}


\begin{document}

\maketitle

\begin{abstract}

Functional magnetic resonance imaging (fMRI) is essential for developing encoding models that identify functional changes in language-related brain areas of individuals with Neurocognitive Disorders (NCD). While large language model (LLM)-based fMRI encoding has shown promise, existing studies predominantly focus on healthy, young adults, overlooking older NCD populations and cognitive level correlations. This paper explores language-related functional changes in older NCD adults using LLM-based fMRI encoding and brain scores, addressing current limitations. We analyze the correlation between brain scores and cognitive scores at both whole-brain and language-related ROI levels. Our findings reveal that higher cognitive abilities correspond to better brain scores, with correlations peaking in the middle temporal gyrus. This study highlights the potential of fMRI encoding models and brain scores for detecting early functional changes in NCD patients.

\end{abstract}

\section{Introduction}

Neurocognitive Disorder (NCD) is a general term for describing neurocognitive decline beyond normal aging caused by various conditions, such as Alzheimer’s disease and brain vascular diseases \cite{blazer2013neurocognitive,lanctot2017apathy}. It poses major challenges to individuals’ well-being and the society \cite{clark2013ncds}. Early detection of NCD is critical because it is possible to halt or even reverse its progression during the early stage, but much less possible at its later stage  \cite{leibing2014earlier,teipel2013relevance}. Deficits in language functions are one of the major symptoms of various types of NCD \cite{appell1982study,gong2022hong}, and the changes of language-related functions in the brain may emerge before structural brain changes and overt NCD symptoms appear \cite{raji2009age}. Therefore, it is promising to detect language-related functional changes in the brain as a mean of early NCD detection. 

Functional magnetic resonance imaging (fMRI) is widely used to study language-related functional changes in the brain. FMRI measures brain activity by recording the blood-oxygenation-level changes in the brain noninvasively with high spatial resolution. And the brain encoding models built upon fMRI signals and large language models (LLMs)  provide researchers in cognitive neuroscience with a powerful computational tool to quantify and locate language-related functions in the human brain, and such models have attracted wide attention in recent years \cite{caucheteux2022brains,gong2023phonemic,caucheteux2023evidence,allen2022massive,tang2023semantic,antonello2024scaling,schrimpf2021neural,caucheteux2022deep,oota2023meg}. They also show us a feasible way to quantify language-related functional changes among older adults and help the early detection of NCD. 


\textbf{The fMRI encoding models}
are used for predicting brain activation from language stimuli. A central goal of such models is to reveal how and where linguistic features at various levels, such as semantic and syntactic, are processed within the brain \cite{gong2023phonemic,caucheteux2023evidence}. The typical pipeline for building up a traditional language-related fMRI encoding model is as follows: 1) A set of linguistic features are extracted first from the same language stimuli that human subjects have heard or read. These features can be expert-designed or -encoded simple features (e.g. word count and spectrum) or higher-level contextual embeddings from layers of an LM; 2) a voxel-wise linear regression is then fitted upon these features to predict the fMRI signals in different voxels or brain regions. Given a fitted encoding model, \textbf{brain scores} will be calculated, i.e. the correlation $r$ or determination coefficient $R^2$ between the predicted and actual fMRI signals. A brain score reflects the strength of association between a brain voxel's activity and a specific language process, depending on which linguistic features were extracted at the very beginning. These brain scores thus can help pinpoint brain voxels and regions involved in the processing of specific language features.

Recent developments in LLMs have made it more efficient to build language-related fMRI encoding models \cite{touvron2023llama,zhang2022opt,achiam2023gpt}. Particularly, the embedding of middle layers of an LLM can effectively represent language features (e.g., semantic and syntactic features) that are highly relevant to NCD symptoms \cite{appell1982study,gong2022hong}, while these features are challenging to code in traditional ways (e.g., manually code) or experimentally controlled. Caucheteux et al. \cite{caucheteux2022brains} have revealed that the similarity between the LLMs and the brain primarily depends on their ability to predict words from context. Based on the encoding model of GPT-2, they further strengthened the role of hierarchical predictive coding in language processing \cite{caucheteux2023evidence}. It is even more encouraging that Tang et al. \cite{tang2023semantic} have demonstrated the feasibility of recovering the intelligible meaning of perceived speech from fMRI signals using a semantic encoding model based on GPT-1. Beyond the GPT family, Antonello et al. \cite{antonello2024scaling} have tested the OPT and LLaMA families and found that the brain score of semantic encoding models increases logarithmically with LLM size. These studies used brain scores obtained from different feature spaces of LLMs to locate brain regions related to language functions of different levels, promoting our understanding of complex language processing in the human brain.

\textbf{Limitations in previous research.} 1) The above-mentioned research on LLM-based fMRI encoding models only involved young and healthy subjects. No study has yet used such models to investigate functional changes in older adults, especially those with NCD. 2) Existing studies mostly use brain scores to identify brain regions relevant to language processing, and only a few studies have reported the relationship between brain scores and subjects’ performance on self-paced reading tasks or story comprehensions \cite{schrimpf2021neural,caucheteux2022deep}. The correlation between brain scores and cognitive levels needs further analysis.

In our study, we built an fMRI encoding model for older adults in the early stage of NCD or at risk and explored the correlation between brain scores and the subjects’ overall cognitive functioning levels. We aim to provide evidence for the feasibility of using brain scores obtained from fMRI encoding models to build interpretable models for the early detection of NCD in the future.
We makes the following contributions:
\begin{itemize}
\item As far as we know, this is the first study that applies the fMRI encoding model based on LlaMA2 to study NCD subjects. The model generates brain scores to quantify the association between brain areas and language functions among NCD subjects;
\item We find that brain scores for the higher cognitive-level group are consistently better than those of the lower cognitive-level group, and the correlation between brain scores and cognition peaks in the middle temporal gyrus ($r$ = $0.368$) and the superior frontal gyrus ($r$ = $0.289$);
\item This study provides a feasible direction for further developing interpretable machine-learning models based on language-related fMRI signals for early NCD detection. 
\end{itemize}

\section{Data Collection}
By the time of analysis, data had been collected from $95$ older adults in the following two tasks. Statistics of subjects are reported in Table~\ref{tab: Dataset}.
\begin{figure}[t]
  
  \centering
  \includegraphics[width=\linewidth]{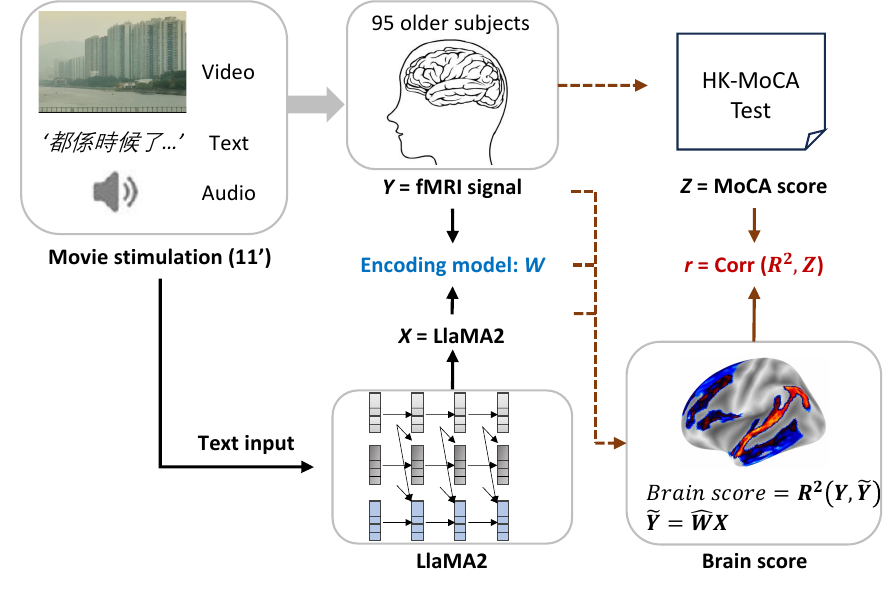}
  \caption{Construction of a language encoding model based on the movie-watching task, and the correlation between brain scores and MoCA scores.}
  
  \label{task_plot}
  \vspace{-1.5em}
\end{figure}

\textbf{Movie-watching fMRI task.} The data in this study was from a group of Hong Kong older adults who were at risk of NCD or had been diagnosed with mild NCD (i.e., mild cognitive impairment). These participants were fMRI scanned when watching an 11-minute clip from a Cantonese movie “Sweet Home.” The movie clip contains everyday scenes with family members engaging in dialogues or monologues. With speech embedded in multimodal information (video, audio, subtitles; see Figure~\ref{task_plot}), this task allows us to examine brain functions involved in the processing of language in a naturalistic context.

The fMRI signals were acquired using a Siemens MAGNETOM Prisma 3 Tesla MRI Scanner with a 64-Channel Head/Neck coil. A multiband (factor = 6) gradient echo echoplanar (EPI) sequence was used to scan the whole brain with the following parameters: repetition time (TR) = 900 ms, echo time (TE) = 24 ms, flip angle = 90°, voxel size = 2 × 2 × 2 mm\textsuperscript{3}, matrix size = 104 × 104, field of view (FoV) = 206 × 206 mm\textsuperscript{2}, and number of slice = 72. For each participant, 736 fMRI images (i.e., 735 TRs) were collected during the movie-watching task. Standardized preprocessing procedures were performed using the SPM12 toolkit  \cite{ashburner2014spm12} for denoising, which includes field map correction, realignment, co-registration, normalization into the standard MNI space, and spatial smoothing. Only fMRI signals from the gray matter in the brain were included for analysis. Head-motion parameters were regressed out as confounders.

\textbf{HK-MoCA test.} The Hong Kong version of the Montreal Cognitive Assessment (HK-MoCA) \cite{wong2009validity} was used to assess participants’ cognitive function. MoCA is a well-established neurocognitive test for NCD diagnosis, which assesses a range of key cognitive functions, such as attention, memory, language, and visuospatial skills \cite{nasreddine2005montreal}. The test score ranges between $0$ and $30$, with a higher score indicating a better cognitive state.

\begin{table}[tbp]
\caption{Statistics of subjects.}
  \label{tab: Dataset}
  \centering
\begin{tabular}{ccccc}
\toprule
\multirow{2}{*}{\textbf{Feature}} & \multicolumn{2}{c}{\textbf{Male} (n = 52)} & \multicolumn{2}{c}{\textbf{Female} (n = 43)} \\ \cline{2-5} 
                                & Mean             & Std.             & Mean               & Std.               \\ \midrule
Age                               & 72.35                & 6.03            & 71.09                  & 6.24               \\
Education                               & 9.25                & 3.90            & 6.70                  & 3.67              \\
MoCA                              &  20.9               & 4.00            & 19.05                  & 4.11               \\ 
\bottomrule
\vspace{-2.5em}
\end{tabular}

\end{table}

\section{Approach}

\subsection{LlaMA2-Cantonese and Context Features}
An open-source LlaMA2-7b-Cantonese model (\url{https://huggingface.co/indiejoseph}) is applied to extract context features for each Cantonese word appearing in the movie. The original LlaMA2-7b, released by Meta \cite{touvron2023llama}, has $32$ layers and is trained on a mix of publicly available online data (with English accounting for $89.7\%$ and Chinese accounting for $0.13\%$), with a context length of $4k$. Since it employs a byte pair encoding algorithm to decompose unknown UTF-8 characters, it can also encode the context knowledge of Cantonese. However, to better adapt to the Cantonese context, the LlaMA2-7b-Cantonese used in this study is further trained on the LlaMA2-7b model using additional Cantonese corpus. 

\begin{figure}[t]
  \centering
  \includegraphics[width=0.85\linewidth]{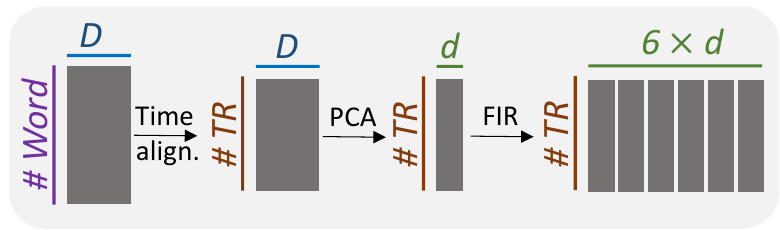}
  \caption{Stimulus matrix construction with time alignment, feature dimension reduction and finite impulse response (FIR) model \cite{caucheteux2023evidence}. D and d are feature dimensions before and after PCA. TR is the repetition time of fMRI.}
  \label{fig:features}
  \vspace{-1em}
\end{figure}

The context feature of the word $s_i$ is extracted based on the next-word-prediction task: for each word-time pair $\left(s_i,t_i \right)$, LlaMA2-Cantonese takes a word sequence $S = \left(s_{i-255}, \cdots, s_{i-1}, s_i\right)$ as input, and its hidden layer activations provide vector embeddings that represent the meaning of $s_i$ within a context length of $256$. This yields a high-dimensional vector-time pair $\left(\boldsymbol{X}_i, t_{i}\right)$ where $\boldsymbol{X}_i$ is a $4096$-dimensional context representation for $s_i$. Then, three steps are conducted before we obtain the final stimulus matrix (as shown in Figure~\ref{fig:features}): (1) these vectors are resampled at times corresponding to the fMRI acquisitions using a three-lobe Lanczos filter \cite{tang2023semantic,deniz2019representation}; (2) vectors are reduced to $d=90$ dimensions using PCA for computational efficiency; (3) finally, based on the linearized finite impulse response (FIR) model \cite{kay2008modeling}, $\left(\boldsymbol{X}_{i-6}, \boldsymbol{X}_{i-5}, \cdots, \boldsymbol{X}_{i-1}\right)$, i.e. context representations from $0.9s$ to $5.4s$ earlier before the timepoint $t_i$, are concatenated to predict the fMRI signal, $\left(y_i,t_i \right)$. We then obtain the final context stimulus matrix $\boldsymbol{X}\in \mathbb{R}^{\#TR\times 6d}$, time-aligned with the processed fMRI signal $\boldsymbol{Y}\in \mathbb{R}^{\#TR\times 1}$ of a brain voxel.

\subsection{Encoding Model and Hyperparameter Selection}

Let $f\left( \boldsymbol{X} \right)$ represent the brain encoding model. Following the work of \cite{antonello2024scaling}, we select $f$ as a voxel-wise linear transformation between $\boldsymbol{X}$ and $\boldsymbol{Y}$ for interpretability. For each subject $s$, voxel $v$, and LlaMA2-Cantonese layer $l_i$, we fit a separate encoding model $f_{s, v}\left(\boldsymbol{X}_{l_i}\right):=\boldsymbol{X}_{l_i} \boldsymbol{W}_{s, v}^{l_i}$, using linearized ridge regression, to predict the fMRI signal $\boldsymbol{Y}_{s,v}$, where $\boldsymbol{W}_{s, v}^{l_i}$ are the learnable weight parameters. The final objective function is:
\begin{equation}
\min_{\boldsymbol{W}_{s,v}^{l_i}}\Vert \boldsymbol{Y}_{s,v}^{l_i}-\boldsymbol{X}_{l_i}\boldsymbol{W}_{s,v}^{l_i} \Vert^2_F+\lambda_{s,v}\Vert \boldsymbol{W}_{s,v}^{l_i} \Vert^2_F
\end{equation}
where $\Vert \cdot \Vert_F$ denotes the Frobenius norm, and $\lambda_{s,v}$ is the only hyperparameter representing the regularization weight. Each $\lambda_{s,v}$ is selected independently for each voxel in each subject.

Specifically, $20\%$ of the data samples (continuous in the time dimension) are held out as the test set. In contrast, the remaining data samples are divided into training and validation sets for regression and hyperparameter selection using a bootstrap method \cite{tang2023semantic}. In each iteration, the model weights are estimated on the training set for each of $10$ possible regularization coefficients (log spaced between $10$ and $1000$). These weights are used to predict responses in the validation set, followed by the calculation of $R^2$ between the actual and predicted fMRI time series. The regularization coefficient for each voxel is chosen based on the value that yields the best performance on the validation set, averaged across 50 bootstraps. After all parameters are finalized, the encoding model is applied to the test set to obtain the brain score (i.e., $R^2$) for each voxel.

\subsection{Experimental Design}
\textbf{Brain score analysis of the whole brain.} 
We computed the brain score for the entire brain of each subject, averaging across all voxels, with various activation layers of LlaMA2-Cantonese. The Pearson correlation between the brain score and the subject group's MoCA score was subsequently calculated using \textit{pearsonr} in Scipy toolkit of Python. 

\textbf{Brain score analysis within language-related ROIs.}
We focus on language-related ROIs for two reasons: 1) early functional changes in NCD subjects occur in language brain areas; 2) the movie-watching task could activate brain regions related to vision, and the ROI analysis can reduce the influence from other cognitive processes to some extent. 

Consequently, our study uses the brain parcellation from the Destrieux cortical deterministic atlas (dated $2009$) \cite{destrieux2009sulcal} to identify language-related ROIs \cite{huth2016natural,zhang2022probing} in both brain hemispheres, including the precuneus, angular gyrus (AG), inferior temporal gyrus (ITG), middle temporal gyrus (MTG), superior temporal gyrus (STG), superior frontal gyrus (SFG), middle frontal gyrus (MFG), and inferior frontal gyrus (IFG). These ROIs correspond to a total of 26 labels in the Destrieux atlas. We then analyzed the brain scores of each ROI and their Pearson correlations with MoCA scores.

\textbf{Statistical significance}
To assess the statistical significance of the Pearson correlations between brain scores and MoCA scores, we conducted two-tailed t-tests with $p<0.05$. The Benjamini-Hochberg False Discovery Rate (FDR) \cite{benjamini2000adaptive} correction was used to adjust the p-values for multiple comparisons.

\begin{figure}[t]
  \centering
  \includegraphics[width=\linewidth]{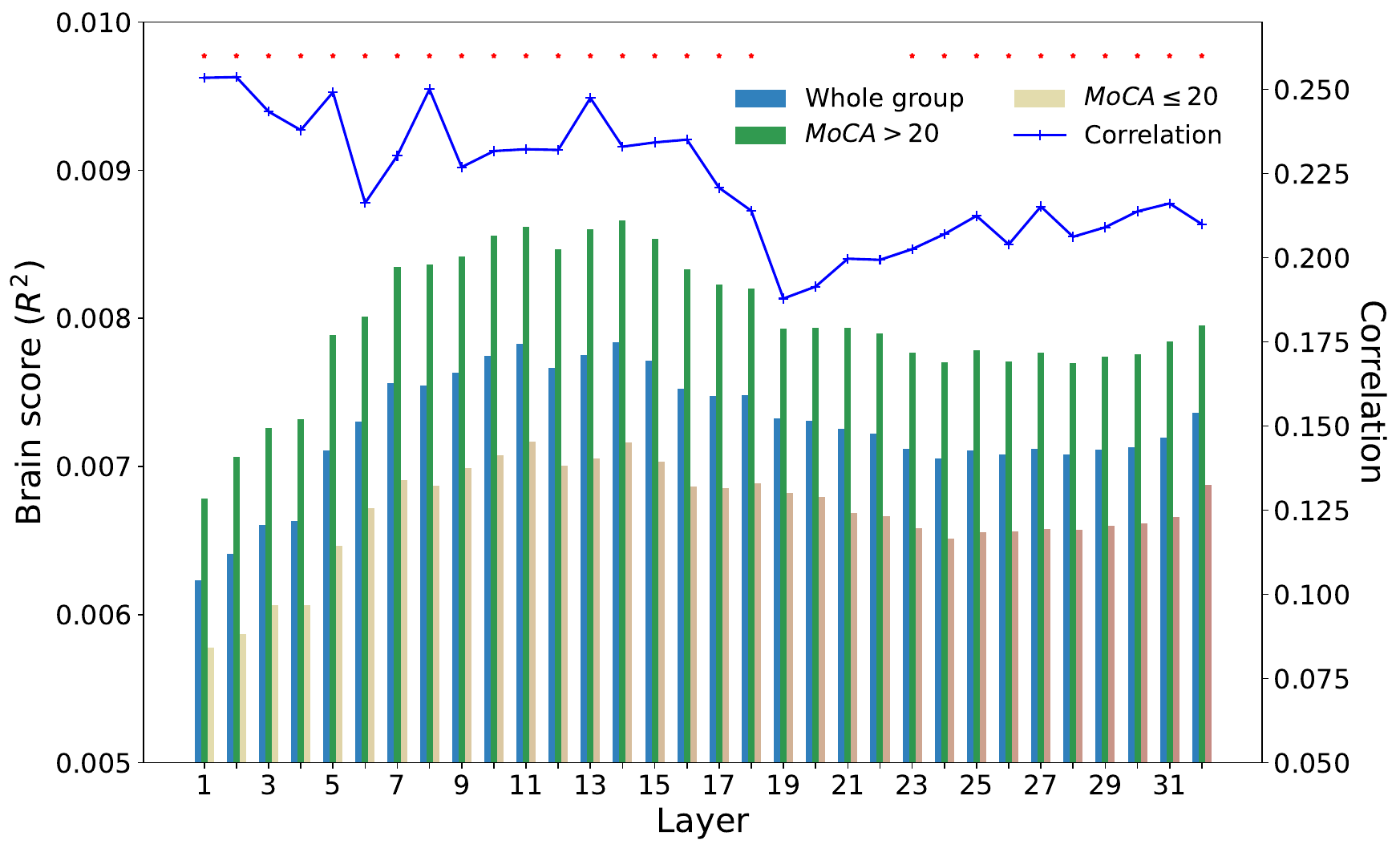}
  \caption{Average brain score across different cognitive groups and activation layers (the bar plot), and the correlation between brain score and MoCA score (the blue line). The red dot means the correlation of that layer is significant.}
  \label{fig:bar}
  \vspace{-0.5em}
\end{figure}

\section{Results and Discussions}

\begin{figure*}[htb]

  \centering
  \includegraphics[width=\textwidth]{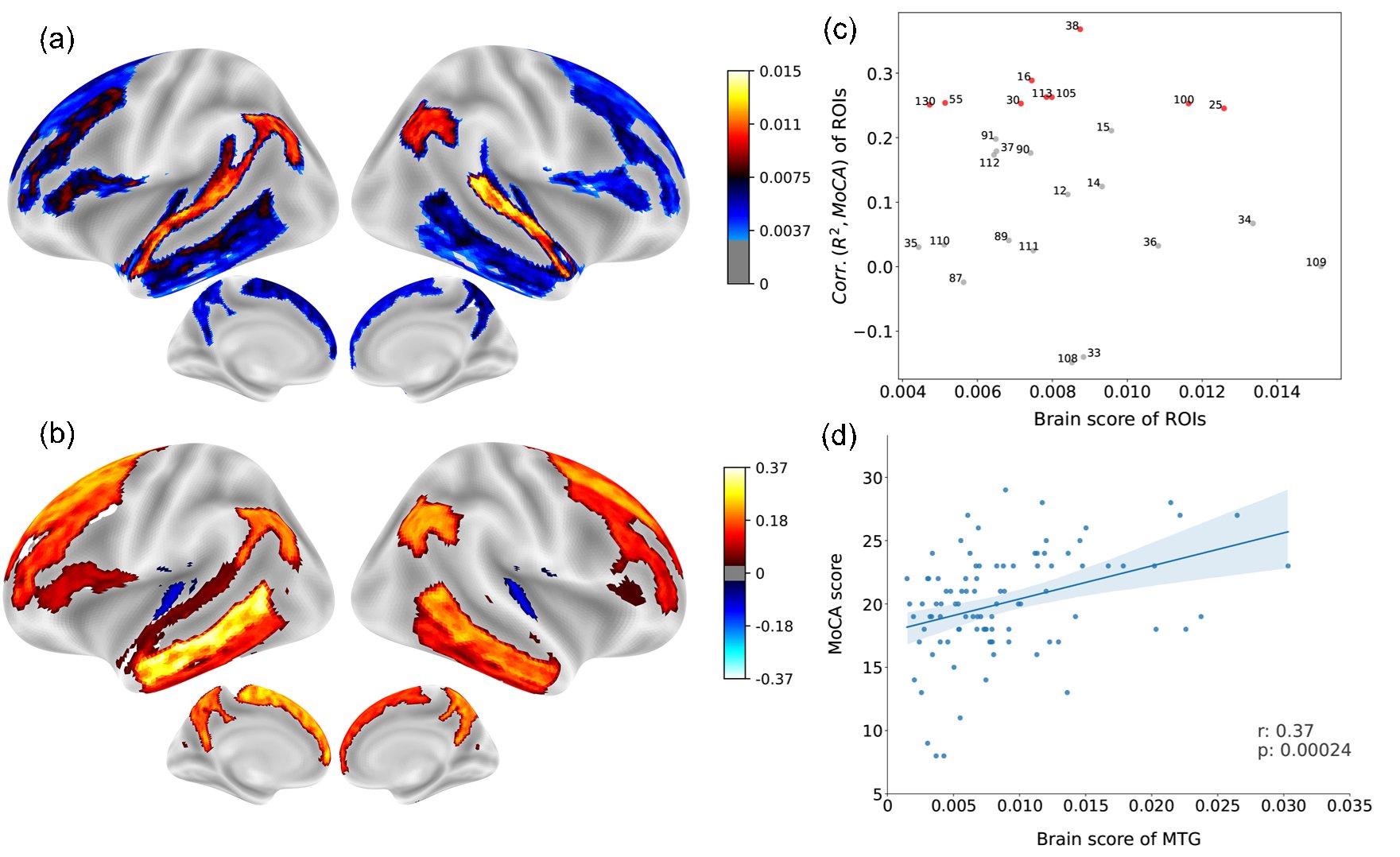}
  
  \caption{(a) Averaged brain score map (activated by the 8th layer) for language-related ROIs. (b) Map of Pearson correlation coefficients between the averaged brain scores and MoCA scores. (c) Relationship between the averaged brain scores of ROIs and correlation coefficients. The number next to the point indicates the label number of the ROI in the Destrieux atlas. Red points represent ROIs with an adjusted $p$-value $< 0.05$; (d) Regression plot of MoCA scores and brain scores for the middle temporal gyrus (MTG).}
  \vspace{-1em}
  \label{rois}
\end{figure*}

\subsection{Brain Score Analysis of the Whole Brain}
The blue bars in Figure~\ref{fig:bar} display the average brain scores for the entire subject group. The $95$ subjects are further divided into two subgroups: the higher cognitive subgroup with MoCA scores $> 20$ and the lower cognitive subgroup with MoCA $\leq 20$, where $20$ is the median of their MoCA scores.


The overall trend indicates that the average brain scores for different cognitive groups reach peaks simultaneously in the relatively early layers (from the 7th to the 16th layer), followed by a gradual decline. Their trend lines align well with \cite{antonello2024scaling}, which reported that the performances of semantic encoding models for three young adults fitted by LlaMA2-30B or LlaMA2-65B reach a peak between $20\%$ and $40\%$ of the layer depth. These similar patterns across different cognitive subgroups and language stimuli suggest both the effectiveness and the general nature of the encoding model fitted by LlaMA2.

Comparing the two cognitive subgroups, the average brain score for the higher cognitive group is consistently better than that of the lower cognitive group. The averaged $\Delta(R^2) = \num{1.28e-3}$ across $32$ layers, with a standard deviation of \num{1.53e-4} indicating that the fMRI signal within the higher cognitive subgroup more closely aligns with context representations. This gap could arise from two possible reasons: 1) individuals with higher cognitive abilities may have followed the dialogue more closely during the task; or 2) their brains could process context information more similarly to LlaMA2.

The Pearson correlation between the brain scores and MoCA scores for the entire group is depicted as the blue line in Figure~\ref{fig:bar}. Correlation coefficients reach $0.25$ in layers $1$, $2$, and $8$, and drop significantly after layer $16$. Along with brain score performance, we focus on the encoding model fitted by the 8th layer in the subsequent ROI analysis.

\subsection{Brain Score Analysis in Language-related ROIs}
In Figure~\ref{rois}, the average brain score and correlation coefficient of each ROI are projected onto the \textit{fsaverage5} brain surface using \textit{vol\underline{~}to\underline{~}surf} in nilearn toolkit. As illustrated in Figure~\ref{rois}c, only $9$ out of $26$ ROIs pass the significance test for correlation ($p$ $< 0.05$ adjusted using FDR): MTG (labels $38$, $113$), Precuneus (labels $30$, $105$), AG (labels $25$, $100$), Left-SFG (label $16$), and SFS (labels $55$, $130$).

The lateral STG (labels $109$, $34$), AG, and the plan-tempo STG in the left hemisphere (label $36$)  exhibit higher brain scores (figure~\ref{rois}a). However, only the AG is significantly correlated with the MoCA score, with $r_{100} = 0.253$ and $r_{25} = 0.245$ (figure~\ref{rois}b,c).  We propose that the STG primarily encodes lower-level semantic information, as supported by \cite{vaidya2022self,millet2022toward}. Consequently, the STG may be associated with perception but not strongly correlated with high-level semantic understanding.

To our surprise,  although the MTG in the left hemisphere has a medium brain score of $\num{8.7e-3}$, it attains the highest correlation coefficient of $r_{38} = 0.37$ (adjusted $p< 0.001$). The detailed regression between the brain score of MTG and the MoCA score is plotted in Figure~\ref{rois}d. These findings are consistent with \cite{caucheteux2022deep}, which reported that the correlation between brain scores and comprehension scores peaks in the AG and MTG.

\subsection{Effect of Cantonese Pretraining}
To investigate the impact of additional Cantonese corpus on LLaMa2-Cantonese, we refitted the encoding model using the representations of the original LLaMa2 and calculated the average brain scores along with their correlations to MoCA scores.
Results in Table~\ref{tab:ablation1} indicate that no significant difference in brain scores or correlations exist, except for the correlation at layer $8$, which is $4.9\%$ higher after Cantonese pretraining. 
This can be attributed to the fact that Cantonese pretraining did not alter the structure or embedding representations of LLaMA2.


\begin{table}[th]
\caption{Performance comparison of LlaMA2-7b and LlaMA2-7b-Cantonese. Corr. refers to Pearson correlation coefficients.}
  \label{tab:ablation1}
  \centering
\begin{tabular}{ccccc}
\toprule

\multirow{2}{*}{\textbf{Layer}} & \multicolumn{2}{c}{\textbf{LlaMA2}} & \multicolumn{2}{c}{\textbf{LlaMA2-Cantonese}} \\ \cline{2-5} 

                                & Brain score             & Corr.             & Brain score               & Corr.                \\ \midrule
1                               & $\num{6.25e-3}$                & 0.2532           & $\num{6.23e-3}$                  & 0.2535               \\
8                               & $\num{7.50e-3}$                & 0.2384            & $\num{7.55e-3}$                  & 0.2500              \\
16                              & $\num{7.53e-3}$                & 0.2331           & $\num{7.53e-3}$                  & 0.2351               \\
24                              & $\num{7.18e-3}$                & 0.1993            & $\num{7.05e-3}$                  & 0.2070               \\
32                              & $\num{7.34e-3}$                & 0.2027            & $\num{7.36e-3}$                  & 0.2100              \\ \bottomrule
   
\end{tabular}
\vspace{-1.5em}
\end{table}

\section{Conclusion}

Focusing on older adults with NCD, this study extracted context representation from LlaMA2-Cantonese to model language stimuli in a movie-watching task and innovatively employed the fMRI encoding model and brain scores to assess their language function. We found that subjects with better cognitive states have significantly higher brain scores, and the correlation pattern peaks at language-related ROIs, e.g. MTG, SFG, and AG. 

The primary limitation of this study is the uncertainty surrounding the extent of semantic or syntactic information contained in the embeddings of LLaMA2-Cantonese. Moreover, the brain areas responsible for language processing may also be activated by semantic stimuli generated through vision \cite{tang2024brain}.
In the future, multi-modal semantic information should be comprehensively considered to construct a robust encoding model for the understanding of the interplay between different modalities and language functions.

\section{Acknowledgements} 
This research is partially supported by the HKSARG Research Grants Council’s Theme-based Research Grant Scheme (Project No. T45-407/19N).

\bibliographystyle{IEEEtran}
\bibliography{mybib}

\begin{thebibliography}{10}
\providecommand{\url}[1]{#1}
\csname url@samestyle\endcsname
\providecommand{\newblock}{\relax}
\providecommand{\bibinfo}[2]{#2}
\providecommand{\BIBentrySTDinterwordspacing}{\spaceskip=0pt\relax}
\providecommand{\BIBentryALTinterwordstretchfactor}{4}
\providecommand{\BIBentryALTinterwordspacing}{\spaceskip=\fontdimen2\font plus
\BIBentryALTinterwordstretchfactor\fontdimen3\font minus \fontdimen4\font\relax}
\providecommand{\BIBforeignlanguage}[2]{{%
\expandafter\ifx\csname l@#1\endcsname\relax
\typeout{** WARNING: IEEEtran.bst: No hyphenation pattern has been}%
\typeout{** loaded for the language `#1'. Using the pattern for}%
\typeout{** the default language instead.}%
\else
\language=\csname l@#1\endcsname
\fi
#2}}
\providecommand{\BIBdecl}{\relax}
\BIBdecl

\bibitem{blazer2013neurocognitive}
D.~Blazer, ``Neurocognitive disorders in dsm-5,'' \emph{American Journal of Psychiatry}, vol. 170, no.~6, pp. 585--587, 2013.

\bibitem{lanctot2017apathy}
K.~L. Lanct{\^o}t, L.~Ag{\"u}era-Ortiz, H.~Brodaty, P.~T. Francis, Y.~E. Geda, Z.~Ismail, G.~A. Marshall, M.~E. Mortby, C.~U. Onyike, P.~R. Padala \emph{et~al.}, ``Apathy associated with neurocognitive disorders: recent progress and future directions,'' \emph{Alzheimer's \& Dementia}, vol.~13, no.~1, pp. 84--100, 2017.

\bibitem{clark2013ncds}
H.~Clark, ``Ncds: a challenge to sustainable human development,'' \emph{The Lancet}, vol. 381, no. 9866, pp. 510--511, 2013.

\bibitem{leibing2014earlier}
A.~Leibing, ``The earlier the better: Alzheimer’s prevention, early detection, and the quest for pharmacological interventions,'' \emph{Culture, Medicine, and Psychiatry}, vol.~38, pp. 217--236, 2014.

\bibitem{teipel2013relevance}
S.~J. Teipel, M.~Grothe, S.~Lista, N.~Toschi, F.~G. Garaci, and H.~Hampel, ``Relevance of magnetic resonance imaging for early detection and diagnosis of alzheimer disease,'' \emph{Medical Clinics}, vol.~97, no.~3, pp. 399--424, 2013.

\bibitem{appell1982study}
J.~Appell, A.~Kertesz, and M.~Fisman, ``A study of language functioning in alzheimer patients,'' \emph{Brain and language}, vol.~17, no.~1, pp. 73--91, 1982.

\bibitem{gong2022hong}
X.~Gong, P.~C. Wong, H.~H. Fung, V.~C. Mok, T.~C. Kwok, J.~Woo, K.~H. Wong, and H.~Meng, ``The hong kong grocery shopping dialog task (hk-gsdt): A quick screening test for neurocognitive disorders,'' \emph{International Journal of Environmental Research and Public Health}, vol.~19, no.~20, p. 13302, 2022.

\bibitem{raji2009age}
C.~A. Raji, O.~Lopez, L.~Kuller, O.~Carmichael, and J.~Becker, ``Age, alzheimer disease, and brain structure,'' \emph{Neurology}, vol.~73, no.~22, pp. 1899--1905, 2009.

\bibitem{caucheteux2022brains}
C.~Caucheteux and J.-R. King, ``Brains and algorithms partially converge in natural language processing,'' \emph{Communications biology}, vol.~5, no.~1, p. 134, 2022.

\bibitem{gong2023phonemic}
X.~L. Gong, A.~G. Huth, F.~Deniz, K.~Johnson, J.~L. Gallant, and F.~E. Theunissen, ``Phonemic segmentation of narrative speech in human cerebral cortex,'' \emph{Nature communications}, vol.~14, no.~1, p. 4309, 2023.

\bibitem{caucheteux2023evidence}
C.~Caucheteux, A.~Gramfort, and J.-R. King, ``Evidence of a predictive coding hierarchy in the human brain listening to speech,'' \emph{Nature human behaviour}, vol.~7, no.~3, pp. 430--441, 2023.

\bibitem{allen2022massive}
E.~J. Allen, G.~St-Yves, Y.~Wu, J.~L. Breedlove, J.~S. Prince, L.~T. Dowdle, M.~Nau, B.~Caron, F.~Pestilli, I.~Charest \emph{et~al.}, ``A massive 7t fmri dataset to bridge cognitive neuroscience and artificial intelligence,'' \emph{Nature neuroscience}, vol.~25, no.~1, pp. 116--126, 2022.

\bibitem{tang2023semantic}
J.~Tang, A.~LeBel, S.~Jain, and A.~G. Huth, ``Semantic reconstruction of continuous language from non-invasive brain recordings,'' \emph{Nature Neuroscience}, vol.~26, no.~5, pp. 858--866, 2023.

\bibitem{antonello2024scaling}
R.~Antonello, A.~Vaidya, and A.~Huth, ``Scaling laws for language encoding models in fmri,'' \emph{Advances in Neural Information Processing Systems}, vol.~36, 2024.

\bibitem{schrimpf2021neural}
M.~Schrimpf, I.~A. Blank, G.~Tuckute, C.~Kauf, E.~A. Hosseini, N.~Kanwisher, J.~B. Tenenbaum, and E.~Fedorenko, ``The neural architecture of language: Integrative modeling converges on predictive processing,'' \emph{Proceedings of the National Academy of Sciences}, vol. 118, no.~45, p. e2105646118, 2021.

\bibitem{caucheteux2022deep}
C.~Caucheteux, A.~Gramfort, and J.-R. King, ``Deep language algorithms predict semantic comprehension from brain activity,'' \emph{Scientific reports}, vol.~12, no.~1, p. 16327, 2022.

\bibitem{oota2023meg}
S.~R. Oota, N.~Trouvain, F.~Alexandre, and X.~Hinaut, ``Meg encoding using word context semantics in listening stories,'' in \emph{INTERSPEECH 2023-24th INTERSPEECH Conference}, 2023.

\bibitem{touvron2023llama}
H.~Touvron, L.~Martin, K.~Stone, P.~Albert, A.~Almahairi, Y.~Babaei, N.~Bashlykov, S.~Batra, P.~Bhargava, S.~Bhosale \emph{et~al.}, ``Llama 2: Open foundation and fine-tuned chat models,'' \emph{arXiv preprint arXiv:2307.09288}, 2023.

\bibitem{zhang2022opt}
S.~Zhang, S.~Roller, N.~Goyal, M.~Artetxe, M.~Chen, S.~Chen, C.~Dewan, M.~Diab, X.~Li, X.~V. Lin \emph{et~al.}, ``Opt: Open pre-trained transformer language models,'' \emph{arXiv preprint arXiv:2205.01068}, 2022.

\bibitem{achiam2023gpt}
J.~Achiam, S.~Adler, S.~Agarwal, L.~Ahmad, I.~Akkaya, F.~L. Aleman, D.~Almeida, J.~Altenschmidt, S.~Altman, S.~Anadkat \emph{et~al.}, ``Gpt-4 technical report,'' \emph{arXiv preprint arXiv:2303.08774}, 2023.

\bibitem{ashburner2014spm12}
J.~Ashburner, G.~Barnes, C.-C. Chen, J.~Daunizeau, G.~Flandin, K.~Friston, S.~Kiebel, J.~Kilner, V.~Litvak, R.~Moran \emph{et~al.}, ``Spm12 manual,'' \emph{Wellcome Trust Centre for Neuroimaging, London, UK}, vol. 2464, no.~4, 2014.

\bibitem{wong2009validity}
A.~Wong, Y.~Y. Xiong, P.~W. Kwan, A.~Y. Chan, W.~W. Lam, K.~Wang, W.~C. Chu, D.~L. Nyenhuis, Z.~Nasreddine, L.~K. Wong \emph{et~al.}, ``The validity, reliability and clinical utility of the hong kong montreal cognitive assessment (hk-moca) in patients with cerebral small vessel disease,'' \emph{Dementia and geriatric cognitive disorders}, vol.~28, no.~1, pp. 81--87, 2009.

\bibitem{nasreddine2005montreal}
Z.~S. Nasreddine, N.~A. Phillips, V.~B{\'e}dirian, S.~Charbonneau, V.~Whitehead, I.~Collin, J.~L. Cummings, and H.~Chertkow, ``The montreal cognitive assessment, moca: a brief screening tool for mild cognitive impairment,'' \emph{Journal of the American Geriatrics Society}, vol.~53, no.~4, pp. 695--699, 2005.

\bibitem{deniz2019representation}
F.~Deniz, A.~O. Nunez-Elizalde, A.~G. Huth, and J.~L. Gallant, ``The representation of semantic information across human cerebral cortex during listening versus reading is invariant to stimulus modality,'' \emph{Journal of Neuroscience}, vol.~39, no.~39, pp. 7722--7736, 2019.

\bibitem{kay2008modeling}
K.~N. Kay, S.~V. David, R.~J. Prenger, K.~A. Hansen, and J.~L. Gallant, ``Modeling low-frequency fluctuation and hemodynamic response timecourse in event-related fmri,'' Wiley Online Library, Tech. Rep., 2008.

\bibitem{destrieux2009sulcal}
C.~Destrieux, B.~Fischl, A.~Dale, and E.~Halgren, ``A sulcal depth-based anatomical parcellation of the cerebral cortex.'' \emph{NeuroImage}, vol.~47, p. S151, 2009.

\bibitem{huth2016natural}
A.~G. Huth, W.~A. De~Heer, T.~L. Griffiths, F.~E. Theunissen, and J.~L. Gallant, ``Natural speech reveals the semantic maps that tile human cerebral cortex,'' \emph{Nature}, vol. 532, no. 7600, pp. 453--458, 2016.

\bibitem{zhang2022probing}
X.~Zhang, S.~Wang, N.~Lin, J.~Zhang, and C.~Zong, ``Probing word syntactic representations in the brain by a feature elimination method,'' in \emph{Proceedings of the AAAI Conference on Artificial Intelligence}, vol.~36, no.~10, 2022, pp. 11\,721--11\,729.

\bibitem{benjamini2000adaptive}
Y.~Benjamini and Y.~Hochberg, ``On the adaptive control of the false discovery rate in multiple testing with independent statistics,'' \emph{Journal of educational and Behavioral Statistics}, vol.~25, no.~1, pp. 60--83, 2000.

\bibitem{vaidya2022self}
A.~R. Vaidya, S.~Jain, and A.~Huth, ``Self-supervised models of audio effectively explain human cortical responses to speech,'' in \emph{International Conference on Machine Learning}.\hskip 1em plus 0.5em minus 0.4em\relax PMLR, 2022, pp. 21\,927--21\,944.

\bibitem{millet2022toward}
J.~Millet, C.~Caucheteux, Y.~Boubenec, A.~Gramfort, E.~Dunbar, C.~Pallier, J.-R. King \emph{et~al.}, ``Toward a realistic model of speech processing in the brain with self-supervised learning,'' \emph{Advances in Neural Information Processing Systems}, vol.~35, pp. 33\,428--33\,443, 2022.

\bibitem{tang2024brain}
J.~Tang, M.~Du, V.~Vo, V.~Lal, and A.~Huth, ``Brain encoding models based on multimodal transformers can transfer across language and vision,'' \emph{Advances in Neural Information Processing Systems}, vol.~36, 2024.

\end{thebibliography}

\end{document}